\documentstyle[prd,aps,preprint,tighten,epsfig]{revtex}

\begin{document}

\draft

\title{Reconstruction of the Neutrino Mixing Matrix and Leptonic
Unitarity Triangles from Long-baseline Neutrino Oscillations}
\author{{\bf Zhi-zhong Xing} ~ and ~ {\bf He Zhang}}
\address{CCAST (World Laboratory), P.O. Box 8730, Beijing 100080, China \\
and Institute of High Energy Physics, Chinese Academy of Sciences, \\
P.O. Box 918 (4), Beijing 100049, China
\footnote{Mailing address} \\
({\it Electronic address: xingzz@mail.ihep.ac.cn,
zhanghe@mail.ihep.ac.cn}) } \maketitle

\begin{abstract}
We derive a new set of sum rules between the neutrino mass and
mixing parameters in vacuum and their effective counterparts in
matter. The moduli of nine genuine lepton mixing matrix elements
can then be calculated in terms of the matter-corrected ones, and
the latter can directly be determined from a variety of
long-baseline neutrino oscillations. We show that it is possible
to reconstruct the leptonic unitarity triangles and CP violation
in a similar parametrization-independent way.
\end{abstract}


\newpage

\section{Introduction}

Thanks to the Super-Kamiokande \cite{SK} and SNO \cite{SNO}
experiments, both solar and atmospheric neutrino oscillations are
convincingly established. It turns out that neutrinos {\it do}
have masses and lepton flavors are really mixed, just as expected
in some grand unified theories. Two neutrino mass-squared
differences and three lepton mixing angles have been measured or
constrained by current neutrino oscillation data
\cite{SK,SNO,KM,CHOOZ,K2K}. A more precise determination of these
parameters has to rely on the new generation of accelerator
neutrino experiments with very long baselines \cite{Long}, in
which leptonic CP violation may also be observed. The terrestrial
matter effects in all long-baseline neutrino experiments must be
taken into account, since they can unavoidably modify the genuine
behaviors of neutrino oscillations in vacuum.

To formulate the probabilities of neutrino oscillations in matter
in the same form as those in vacuum, one may define the {\it
effective} neutrino masses $\tilde{m}_i$ and the {\it effective}
lepton flavor mixing matrix $\tilde{V}$ in which the terrestrial
matter effects are already included. In this common approach, it
is necessary to find out the relationship between the fundamental
quantities of neutrino mixing in vacuum ($m_i$ and $V$) and their
effective counterparts in matter ($\tilde{m}_i$ and $\tilde{V}$).
The exact formulas of $\tilde{m}_i$ and $\tilde{V}$ as functions
of $m_i$ and $V$ have been achieved by a number of authors
\cite{Barger,Zaglauer,Xing00,Zhang05}, and the similar expressions
of $m_i$ and $V$ in terms of $\tilde{m}_i$ and $\tilde{V}$ have
been derived in Ref. \cite{Xing01}. The latter case is
equivalently interesting in phenomenology, because our physical
purpose is to determine the fundamental parameters of lepton
flavor mixing from the effective ones, whose values can directly
be measured from a variety of long-baseline neutrino oscillation
experiments.

This paper aims to reconstruct the genuine neutrino mixing matrix
$V$ and its unitarity triangles from possible long-baseline
neutrino oscillations with terrestrial matter effects. Our work is
remarkably different from the existing ones
\cite{Barger,Zaglauer,Xing00,Zhang05,Xing01} in the following
aspects:
\begin{itemize}
\item We derive a new set of sum rules between $(m_i, V)$ and
$(\tilde{m}_i, \tilde{V})$. It may perfectly complement the sum
rules obtained in Ref. \cite{SUM}. We show that similar results
can be achieved in the four-neutrino mixing scheme.
\item Our sum
rules allow us to calculate the moduli of $V_{\alpha i}$ (for
$\alpha =e, \mu, \tau$ and $i=1, 2,3$) in terms of those of
$\tilde{V}_{\alpha i}$. This approach, which results in some much
simpler relations between $|V_{\alpha i}|^2$ and
$|\tilde{V}_{\alpha i}|^2$, proves to be more useful than that
proposed in our previous work \cite{Zhang05}. The sides of three
unitarity triangles can also be derived in a similar way.
\item We
express the neutrino oscillation probabilities in terms of
$|\tilde{V}_{\alpha i}|^2$ and the matter-corrected Jarlskog
parameter $\tilde{J}$ \cite{J}. The fundamental lepton flavor
mixing matrix $V$ and its unitarity triangles can then be
reconstructed from possible long-baseline neutrino oscillations
straightforwardly and parametrization-independently.
\end{itemize}
The only assumption to be made is a constant earth density
profile. Such an assumption is rather reasonable and close to
reality for most of the presently-proposed terrestrial
long-baseline neutrino oscillation experiments \cite{Long}, in
which the neutrino beam is not expected to go through the earth's
core.

The remaining parts of this paper are organized as follows. In
section II, we figure out a new set of sum rules between $(m_i,
V)$ and $(\tilde{m}_i, \tilde{V})$ and discuss its extension in
the four-neutrino mixing scheme. Section III is devoted to the
calculation of $|V_{\alpha i}|^2$ in terms of $|\tilde{V}_{\alpha
i}|^2$. We also show how to derive the sides of three unitarity
triangles in a similar approach. In section IV, the neutrino
oscillation probabilities are presented in terms of
$|\tilde{V}_{\alpha i}|^2$ and $\tilde{J}$. We illustrate the
dependence of different oscillation terms on the neutrino beam
energy and the baseline length. Section V is devoted to a brief
summary of our main results.

\section{New sum rules between $V$ and $\tilde{V}$}

In the basis where the flavor eigenstates of charged leptons are
identified with their mass eigenstates, the lepton flavor mixing
matrix $V$ is defined to link the neutrino mass eigenstates
($\nu_1, \nu_2, \nu_3$) to the neutrino flavor eigenstates
($\nu_e, \nu_\mu, \nu_\tau$):
\begin{equation}
\left ( \matrix{\nu_e \cr \nu_\mu \cr \nu_\tau \cr} \right ) =
\left ( \matrix{ V_{e1} & V_{e2} & V_{e3} \cr V_{\mu 1}   & V_{\mu
2} & V_{\mu 3} \cr V_{\tau 1}  & V_{\tau 2} & V_{\tau 3} \cr}
\right ) \left ( \matrix{\nu_1 \cr \nu_2 \cr \nu_3 \cr} \right )
\; .
\end{equation}
A similar definition can be made for $\tilde{V}$, the effective
counterpart of $V$ in matter. The strength of CP violation in
normal neutrino oscillations is measured by a rephasing-invariant
quantity $J$ (in vacuum) or $\tilde{J}$ (in matter), the so-called
Jarlskog parameter \cite{J}:
\begin{eqnarray}
{\rm Im} \left (V_{\alpha i}V_{\beta j} V^*_{\alpha j}V^*_{\beta i} \right )
& = & J \sum_{\gamma,k} \left (\epsilon^{~}_{\alpha \beta \gamma}
\epsilon^{~}_{ijk} \right ) \; ,
\nonumber \\
{\rm Im} \left (\tilde{V}_{\alpha i}\tilde{V}_{\beta j}
\tilde{V}^*_{\alpha j}\tilde{V}^*_{\beta i} \right )
& = & \tilde{J} \sum_{\gamma,k} \left (\epsilon^{~}_{\alpha \beta \gamma}
\epsilon^{~}_{ijk} \right ) \; ,
\end{eqnarray}
where the Greek subscripts $(\alpha, \beta, \gamma)$ and the Latin
subscripts $(i, j, k)$ run over $(e, \mu, \tau)$ and $(1, 2, 3)$,
respectively.

The effective Hamiltonian responsible for the propagation of neutrinos
in vacuum or in matter can be written as
\begin{eqnarray}
H_\nu & = & \frac{1}{2E} (M_\nu M^\dagger_\nu ) = \frac{1}{2E}
\left (V D^2_\nu V^{\dagger} \right ) \; ,
\nonumber \\
\tilde{H}_\nu & = & \frac{1}{2E} (\tilde{M}_\nu
\tilde{M}^\dagger_\nu ) = \frac{1}{2E} \left (\tilde{V}
\tilde{D}^2_\nu \tilde{V}^{\dagger} \right ) \; ,
\end{eqnarray}
where $D_\nu \equiv {\rm Diag}\{m^2_1, m^2_2, m^2_3 \}$ and
$\tilde{D}_\nu \equiv {\rm Diag}\{\tilde{m}^2_1, \tilde{m}^2_2,
\tilde{m}^2_3 \}$, $E$ is the neutrino beam energy, $M_\nu$ and
$\tilde{M}_\nu$ denote the corresponding genuine and
matter-corrected neutrino mass matrices in the chosen flavor
basis, $m_i$ and $\tilde{m}_i$ (for $i=1, 2, 3$) stand
respectively for the neutrino masses in vacuum and those in
matter. The deviation of $\tilde{H}_\nu$ from $H_\nu$ results
non-trivially from the charged-current contribution to the $\nu_e
e^-$ forward scattering \cite{Wolfenstein}, when neutrinos travel
through a normal material medium like the earth:
\begin{equation}
\tilde{H}_\nu - H_\nu \equiv \frac{1}{2E} \Omega_\nu = {\rm Diag}
\{a, 0, 0 \} \; , ~~
\end{equation}
where $a = \sqrt{2} G_{\rm F} N_e$ with $N_e$ being the background
density of electrons. Subsequently we assume a constant earth
density profile (i.e., $N_e$ = constant), which is a very good
approximation for most of the long-baseline neutrino experiments
proposed at present.

Eq. (3) implies that $(M_\nu M^\dagger_\nu )^n = V D^{2n}_\nu
V^\dagger$ and $(\tilde{M}_\nu \tilde{M}^\dagger_\nu )^n =
\tilde{V} \tilde{D}^{2n}_\nu \tilde{V}^\dagger$ hold, in which
$n=0, \pm 1, \pm 2$, etc. To be explicit, we obtain
\begin{eqnarray}
(M_\nu M^\dagger_\nu )^n_{\alpha \beta} & = & \sum^3_{i=1} \left
(m^{2n}_i V_{\alpha i}V^*_{\beta i} \right ) \; ,
\nonumber \\
(\tilde{M}_\nu \tilde{M}^\dagger_\nu )^n_{\alpha \beta} & = &
\sum^3_{i=1} \left (\tilde{m}^{2n}_i \tilde{V}_{\alpha i}
\tilde{V}^*_{\beta i} \right ) \; ,
\end{eqnarray}
where $\alpha$ and $\beta$ run over $e$, $\mu$ and $\tau$. The
simplest connection between $(M_\nu M^\dagger_\nu )^n$ and
$(\tilde{M}_\nu \tilde{M}^\dagger_\nu )^n$ is their linear
relation with $n=1$; i.e.,
\begin{equation}
\tilde{M}_\nu \tilde{M}^\dagger_\nu = M_\nu M^\dagger_\nu +
\Omega_\nu \; ,
\end{equation}
as indicated by Eqs. (3) and (4). With the help of Eqs. (5) and
(6), one may easily find
\begin{equation}
\sum^3_{i=1} \left (\tilde{m}^2_i \tilde{V}_{\alpha i}
\tilde{V}^*_{\beta i} \right ) = \sum^3_{i=1} \left (m^2_i
V_{\alpha i}V^*_{\beta i} \right ) + A \delta_{\alpha
e} \delta_{e \beta} \; ,
\end{equation}
where $A=2Ea$. In the $\alpha \neq \beta$ case, Eq. (7) reproduces
the sum rules obtained in Ref. \cite{SUM}. A new set of sum rules
can be achieved, if the square relation
\begin{equation}
(\tilde{M}_\nu \tilde{M}^\dagger_\nu)^2 = (M_\nu M^\dagger_\nu)^2
+ (M_\nu M^\dagger_\nu) \Omega_\nu + \Omega_\nu (M_\nu
M^\dagger_\nu) + \Omega^2_\nu
\end{equation}
is taken into account. Resolving Eq. (8) by use of Eqs. (4) and
(5), we arrive at
\begin{equation}
\sum^3_{i=1} \left (\tilde{m}^4_i \tilde{V}_{\alpha i}
\tilde{V}^*_{\beta i} \right ) = \sum^3_{i=1} \left \{m^2_i \left
[m^2_i + A \left (\delta_{\alpha e} + \delta_{e \beta} \right )
\right ] V_{\alpha i}V^*_{\beta i} \right \} + A^2 \delta_{\alpha
e}\delta_{e \beta} \; .
\end{equation}
Eqs. (7) and (9), together with the unitarity conditions
\begin{equation}
\sum^3_{i=1} (\tilde{V}_{\alpha i} \tilde{V}^*_{\beta i} ) =
\sum^3_{i=1} (V_{\alpha i} V^*_{\beta i} ) = \delta_{\alpha \beta}
\; ,
\end{equation}
constitute a full set of linear equations of $V_{\alpha i}
V^*_{\beta i}$ or $\tilde{V}_{\alpha i} \tilde{V}^*_{\beta i}$ for
$i=1,2$ and 3. We shall make use of these equations to derive the
concrete expressions of $V_{\alpha i} V^*_{\beta i}$ in terms of
$m_i$, $\tilde{m}_i$ and $\tilde{V}_{\alpha i} \tilde{V}^*_{\beta
i}$ in the next section.

It is worth remarking that the sum rules in Eq. (9) are completely
different from those presented in Refs.
\cite{Zhang05,Review,Kimura,Scott}, where the relationship between
$H^{-1}_\nu \det H_\nu$ and $\tilde{H}^{-1}_\nu \det
\tilde{H}_\nu$ has been used. Our new result can appreciably
simplify the calculations of $|V_{\alpha i}|^2$ in terms of
$|\tilde{V}_{\alpha i}|^2$ (or vice versa), as one may see later.

Eqs. (7) and (9) may be generalized to include the mixing between
one sterile neutrino ($\nu_s$) and three active neutrinos
($\nu_e$, $\nu_\mu$ and $\nu_\tau$). In this case, $V$ is
redefined to link the neutrino mass eigenstates $(\nu_0, \nu_1,
\nu_2, \nu_3)$ to the neutrino flavor eigenstates $(\nu_s, \nu_e,
\nu_\mu, \nu_\tau)$:
\begin{equation}
\left ( \matrix{\nu_s \cr \nu_e \cr \nu_\mu \cr \nu_\tau \cr}
\right ) = \left ( \matrix{ V_{s0} & V_{s1} & V_{s2} & V_{s3} \cr
V_{e0} & V_{e1} & V_{e2} & V_{e3} \cr V_{\mu 0} & V_{\mu 1} &
V_{\mu 2} & V_{\mu 3} \cr V_{\tau 0} & V_{\tau 1}  & V_{\tau 2} &
V_{\tau 3} \cr} \right ) \left ( \matrix{\nu_0 \cr \nu_1 \cr \nu_2
\cr \nu_3 \cr} \right ) \; ,
\end{equation}
and $\tilde{V}$ can be redefined in the same manner. The matrices
$D_\nu$, $\tilde{D}_\nu$ and $\Omega_\nu$ in Eqs. (3) and (4) are
now rewritten as
\begin{eqnarray}
D_\nu & = & {\rm Diag} \left \{m^2_0, m^2_1, m^2_2, m^2_3 \right
\} \; ,
\nonumber \\
\tilde{D}_\nu & = & {\rm Diag} \left \{\tilde{m}^2_0,
\tilde{m}^2_1, \tilde{m}^2_2, \tilde{m}^2_3 \right \} \; ,
\nonumber \\
\Omega_\nu & = & {\rm Diag} \left \{A', A, 0, 0 \right \} \; ,
\end{eqnarray}
in which $m_0$ denotes the sterile neutrino's mass, and
$A'=\sqrt{2}G_{\rm F}N_nE$ with $N_n$ being the background density
of neutrons \cite{Grimus}. Different from $A$, $A'$ measures the
universal neutral-current interactions of $\nu_e$, $\nu_\mu$ and
$\nu_\tau$ with terrestrial matter. Both $A$ and $A'$ are assumed
to be constant throughout this work. Then two sets of sum rules
can respectively be derived from the linear and square relations
given in Eqs. (6) and (8):
\begin{eqnarray}
\sum^3_{i=0} \left (\tilde{m}^2_i \tilde{V}_{\alpha i}
\tilde{V}^*_{\beta i} \right ) & = & \sum^3_{i=0} \left (m^2_i
V_{\alpha i}V^*_{\beta i} \right ) + A' \delta_{\alpha s}
\delta_{s \beta} + A \delta_{\alpha e} \delta_{e \beta} \; ,
\nonumber \\
\sum^3_{i=0} \left (\tilde{m}^4_i \tilde{V}_{\alpha i}
\tilde{V}^*_{\beta i} \right ) & = & \sum^3_{i=0} \left \{m^2_i
\left [m^2_i + A' \left (\delta_{\alpha s} + \delta_{s \beta}
\right ) + A \left (\delta_{\alpha e} + \delta_{e \beta} \right )
\right ] V_{\alpha i}V^*_{\beta i} \right \}
\nonumber \\
& & + {A'}^2 \delta_{\alpha s}\delta_{s \beta} + A^2
\delta_{\alpha e}\delta_{e \beta} \; .
\end{eqnarray}
Of course, one may also make use of the relation between $(M_\nu
M^\dagger_\nu )^3$ and $(\tilde{M}_\nu \tilde{M}^\dagger_\nu )^3$
to derive another set of sum rules between $(m_i, V)$ and
$(\tilde{m}_i, \tilde{V})$, although the relevant calculations are
somehow lengthy. It is then possible to establish a full set of
linear equations of $\tilde{V}_{\alpha i} \tilde{V}^*_{\beta i}$
or $V_{\alpha i} V^*_{\beta i}$ for $i=0,1,2$ and 3 in the
four-neutrino mixing scheme, from which both the moduli of
$V_{\alpha i}$ and the sides of six unitarity quadrangles of $V$
\cite{Guo} can be derived in terms of $A$, $A'$ and the neutrino
mass and mixing parameters in matter. Such an idea and its
phenomenological consequences will be elaborated elsewhere.

\section{Moduli and unitarity triangles of $V$}

Now let us calculate the moduli of $V_{\alpha i}$ by using Eqs.
(7), (9) and (10) in the conventional three-neutrino mixing
scheme, assuming that $|\tilde{V}_{\alpha i}|^2$ can directly be
determined from possible long-baseline neutrino oscillation
experiments. Those three equations are rewritten, in the $\alpha =
\beta$ case, as follows:
\begin{equation}
O_1 \left ( \matrix{ |V_{\alpha 1}|^2 \cr |V_{\alpha 2}|^2 \cr
|V_{\alpha 3}|^2 \cr} \right ) = \tilde{O}_1 \left ( \matrix{
|\tilde{V}_{\alpha 1}|^2 \cr |\tilde{V}_{\alpha 2}|^2 \cr
|\tilde{V}_{\alpha 3}|^2 \cr} \right ) - \left ( \matrix{ 0 \cr A
\cr A^2 \cr} \right ) \delta_{\alpha e} \; ,
\end{equation}
where
\begin{eqnarray}
O_1 & = & \left ( \matrix{ 1 & 1 & 1 \cr m_1^2 & m_2^2 & m_3^2 \cr
m_1^2 \left (m_1^2 + 2A \delta_{\alpha e} \right ) & m_2^2 \left
(m_2^2 + 2A \delta_{\alpha e} \right ) & m_3^2 \left (m_3^2 + 2A
\delta_{\alpha e} \right ) \cr} \right ) \; ,
\nonumber \\
\tilde{O}_1 & = & \left ( \matrix{ 1 & 1 & 1 \cr \tilde{m}_1^2 &
\tilde{m}_2^2 & \tilde{m}_3^2 \cr \tilde{m}_1^4 & \tilde{m}_2^4 &
\tilde{m}_3^4 \cr} \right ) \; .
\end{eqnarray}
With the help of the relationship \cite{Review}
\begin{equation}
\sum^3_{i=1} \tilde{m}^2_i = \sum^3_{i=1} m^2_i + A \; ,
\end{equation}
we solve Eq. (14) and obtain the following exact results:
\begin{eqnarray}
|{V}_{e1}|^{2} & = &
\frac{\widehat\Delta_{13}\widehat{\Delta}_{12}}
{{\Delta}_{13}{\Delta}_{12}}|\tilde{V}_{e1}|^{2} +
\frac{\widehat\Delta_{13}\widehat{\Delta}_{11}}{{\Delta}_{13}
{\Delta}_{12}}|\tilde{V}_{e2}|^{2}+\frac{\widehat{\Delta}_{11}
\widehat{\Delta}_{12}}{{\Delta}_{13}{\Delta}_{12}}
|\tilde{V}_{e3}|^{2} \ ,
\nonumber \\
|{V}_{e2}|^{2} & = &
\frac{\widehat\Delta_{22}\widehat{\Delta}_{23}}
{{\Delta}_{21}{\Delta}_{23}}|\tilde{V}_{e1}|^{2} +
\frac{\widehat\Delta_{21}\widehat{\Delta}_{23}}{{\Delta}_{21}
{\Delta}_{23}}|\tilde{V}_{e2}|^{2} + \frac{\widehat{\Delta}_{21}
\widehat{\Delta}_{22}}{{\Delta}_{21}{\Delta}_{23}}
|\tilde{V}_{e3}|^{2} \ ,
\nonumber \\
|{V}_{e3}|^{2} & = &
\frac{\widehat\Delta_{32}\widehat{\Delta}_{33}}
{{\Delta}_{32}{\Delta}_{31}}|\tilde{V}_{e1}|^{2} +
\frac{\widehat\Delta_{33}\widehat{\Delta}_{31}}{{\Delta}_{32}
{\Delta}_{31}}|\tilde{V}_{e2}|^{2}+\frac{\widehat{\Delta}_{32}
\widehat{\Delta}_{31}}{{\Delta}_{32}
{\Delta}_{31}}|\tilde{V}_{e3}|^{2} \ ;
\end{eqnarray}
and
\begin{eqnarray}
|{V}_{\mu 1}|^{2} & = &
\frac{\widehat{\Delta}_{21}\widehat{\Delta}_{31}}
{{\Delta}_{21}{\Delta}_{31}}|\tilde{V}_{\mu 1}|^{2} +
\frac{\widehat{\Delta}_{22}\widehat{\Delta}_{32}} {{\Delta}_{21}
{\Delta}_{31}}|\tilde{V}_{\mu 2}|^{2} +
\frac{\widehat{\Delta}_{23} \widehat{\Delta}_{33}} {{\Delta}_{21}
{\Delta}_{31}} |\tilde{V}_{\mu 3}|^{2} \ ,
\nonumber \\
|{V}_{\mu 2}|^{2} & = &
\frac{\widehat{\Delta}_{11}\widehat{\Delta}_{31}}
{{\Delta}_{12}{\Delta}_{32}}|\tilde{V}_{\mu 1}|^{2} +
\frac{\widehat{\Delta}_{12}\widehat{\Delta}_{32}} {{\Delta}_{12}
{\Delta}_{32}}|\tilde{V}_{\mu 2}|^{2} +
\frac{\widehat{\Delta}_{13}\widehat{\Delta}_{33}} {{\Delta}_{12}
{\Delta}_{32}} |\tilde{V}_{\mu 3}|^{2} \ ,
\nonumber \\
|{V}_{\mu 3}|^{2} & = &
\frac{\widehat{\Delta}_{11}\widehat{\Delta}_{21}}
{{\Delta}_{13}{\Delta}_{23}}|\tilde{V}_{\mu 1}|^{2} +
\frac{\widehat{\Delta}_{12}\widehat{\Delta}_{22}}{{\Delta}_{13}
{\Delta}_{23}}|\tilde{V}_{\mu 2}|^{2} +
\frac{\widehat{\Delta}_{13}\widehat{\Delta}_{23}}{{\Delta}_{13}
{\Delta}_{23}}|\tilde{V}_{\mu 3}|^{2} \ ;
\end{eqnarray}
and
\begin{eqnarray}
|{V}_{\tau 1}|^{2} & = &
\frac{\widehat{\Delta}_{21}\widehat{\Delta}_{31}}
{{\Delta}_{21}{\Delta}_{31}}|\tilde{V}_{\tau 1}|^{2} +
\frac{\widehat{\Delta}_{22}\widehat{\Delta}_{32}}{{\Delta}_{21}
{\Delta}_{31}}|\tilde{V}_{\tau
2}|^{2}+\frac{\widehat{\Delta}_{23}\widehat{\Delta}_{33}}
{{\Delta}_{21} {\Delta}_{31}} |\tilde{V}_{\tau 3}|^{2} \ ,
\nonumber \\
|{V}_{\tau 2}|^{2} & = &
\frac{\widehat{\Delta}_{11}\widehat{\Delta}_{31}}
{{\Delta}_{12}{\Delta}_{32}}|\tilde{V}_{\tau 1}|^{2} +
\frac{\widehat{\Delta}_{12}\widehat{\Delta}_{23}}{{\Delta}_{12}
{\Delta}_{23}}|\tilde{V}_{\tau 2}|^{2} +
\frac{\widehat{\Delta}_{13}\widehat{\Delta}_{33}} {{\Delta}_{12}
{\Delta}_{32}}|\tilde{V}_{\tau 3}|^{2} \ ,
\nonumber \\
|{V}_{\tau 3}|^{2} & = &
\frac{\widehat{\Delta}_{11}\widehat{\Delta}_{21}}
{{\Delta}_{13}{\Delta}_{23}}|\tilde{V}_{\tau 1}|^{2} +
\frac{\widehat{\Delta}_{12}\widehat{\Delta}_{22}}{{\Delta}_{13}
{\Delta}_{23}}|\tilde{V}_{\tau 2}|^{2} +
\frac{\widehat{\Delta}_{13}\widehat{\Delta}_{23}} {{\Delta}_{13}
{\Delta}_{23}}|\tilde{V}_{\tau 3}|^{2} \ ,
\end{eqnarray}
in which the neutrino mass-squared differences
$\widehat{\Delta}_{ij} \equiv m^2_i - \tilde{m}^2_j$ and
$\widetilde{\Delta}_{ij} \equiv \tilde{m}^2_i - \tilde{m}^2_j$ are
defined. One can see that these results are more instructive and
much simpler than those obtained in Ref. \cite{Zhang05}, because
we have introduced a new set of sum rules in Eq. (9).

Next we calculate $V_{\alpha i} V^*_{\beta i}$ in terms of
$\tilde{V}_{\alpha i} \tilde{V}^*_{\beta i}$ (for $\alpha \neq
\beta$). The former can form three unitarity triangles in the
complex plane, which were originally named as $\triangle_e$ with
$(\alpha, \beta) =(\mu, \tau)$, $\triangle_\mu$ with $(\alpha,
\beta) = (\tau, e)$, and $\triangle_\tau$ with $(\alpha, \beta) =
(e, \mu)$ in Ref. \cite{FX}. Their effective counterparts in
matter are then referred to as $\widetilde{\triangle}_e$,
$\widetilde{\triangle}_\mu$ and $\widetilde{\triangle}_\tau$.
Taking account of Eqs. (7), (9) and (10), one may easily write
down a full set of equations of $V_{\alpha i} V^*_{\beta i}$ with
$\alpha \neq \beta$:
\begin{equation}
O_2 \left ( \matrix{ V_{\alpha 1} V^*_{\beta 1} \cr V_{\alpha 2}
V^*_{\beta 2} \cr V_{\alpha 3} V^*_{\beta 3} \cr} \right ) =
\tilde{O}_2 \left ( \matrix{ \tilde{V}_{\alpha 1}
\tilde{V}^*_{\beta 1} \cr \tilde{V}_{\alpha 2} \tilde{V}^*_{\beta
2} \cr \tilde{V}_{\alpha 3}\tilde{V}^*_{\beta 3} \cr}\right ) \; ,
\end{equation}
where
\begin{eqnarray}
O_2 & = & \left ( \matrix{ 1 & 1 & 1 \cr m_1^2 & m_2^2 & m_3^2 \cr
m_1^2 \left [m_1^2 + A \left (\delta_{\alpha e} + \delta_{e \beta}
\right ) \right ] & m_2^2 \left [m_2^2 + A \left (\delta_{\alpha
e} + \delta_{e \beta} \right ) \right ] & m_3^2 \left [m_3^2 + A
\left (\delta_{\alpha e} + \delta_{e \beta} \right ) \right ] \cr}
\right ) \; ,
\nonumber \\
\tilde{O}_2 & = & \left ( \matrix{ 1 & 1 & 1 \cr \tilde{m}_1^2 &
\tilde{m}_2^2 & \tilde{m}_3^2 \cr \tilde{m}_1^4 & \tilde{m}_2^4 &
\tilde{m}_3^4 \cr} \right ) \; .
\end{eqnarray}
We solve Eq. (20) and arrive at
\begin{eqnarray}
{V}_{\mu 1}{V}_{\tau 1}^{\ast} & = &
\frac{(\widehat{\Delta}_{21}+\widehat{\Delta}_{33})
\widetilde\Delta_{31}}{{\Delta}_{21} {\Delta}_{31}}\tilde{V}_{\mu
1}\tilde{V}_{\tau 1}^{\ast}+
\frac{(\widehat{\Delta}_{22}+\widehat{\Delta}_{33})
\widetilde\Delta_{32}}{{\Delta}_{21} {\Delta}_{31}}\tilde{V}_{\mu
2}\tilde{V}_{\tau 2}^{\ast} \ ,
\nonumber  \\
{V}_{\mu 2}{V}_{\tau 2}^{\ast} & = &
\frac{(\widehat{\Delta}_{32}+\widehat{\Delta}_{11})
\widetilde\Delta_{21}}{{\Delta}_{32} {\Delta}_{21}}\tilde{V}_{\mu
2}\tilde{V}_{\tau 2}^{\ast}+
\frac{(\widehat{\Delta}_{11}+\widehat{\Delta}_{33})
\widetilde\Delta_{31}}{{\Delta}_{32} {\Delta}_{21}}\tilde{V}_{\mu
3}\tilde{V}_{\tau 3}^{\ast} \ ,
\nonumber  \\
{V}_{\mu 3}{V}_{\tau 3}^{\ast} & = &
\frac{(\widehat{\Delta}_{13}+\widehat{\Delta}_{22})
\widetilde\Delta_{23}}{{\Delta}_{13} {\Delta}_{23}}\tilde{V}_{\mu
3}\tilde{V}_{\tau 3}^{\ast}+
\frac{(\widehat{\Delta}_{11}+\widehat{\Delta}_{22})
\widetilde\Delta_{21}}{{\Delta}_{13} {\Delta}_{23}} \tilde{V}_{\mu
1}\tilde{V}_{\tau 1}^{\ast} \
\end{eqnarray}
for $\triangle_e$; and
\begin{eqnarray}
{V}_{\tau 1}{V}_{e 1}^{\ast} & = &
\frac{\widehat{\Delta}_{12}\widetilde\Delta_{31}}{{\Delta}_{12}
{\Delta}_{31}}\tilde{V}_{\tau 1}\tilde{V}_{e 1}^{\ast} +
\frac{\widehat{\Delta}_{11}
\widetilde\Delta_{32}}{{\Delta}_{12}{\Delta}_{31}} \tilde{V}_{\tau
2}\tilde{V}_{e 2}^{\ast} \ ,
\nonumber  \\
{V}_{\tau 2}{V}_{e 2}^{\ast} & = &
\frac{\widehat{\Delta}_{23}\widetilde\Delta_{21}}{{\Delta}_{23}
{\Delta}_{21}}\tilde{V}_{\tau 2}\tilde{V}_{e 2}^{\ast} +
\frac{\widehat{\Delta}_{22}
\widetilde\Delta_{31}}{{\Delta}_{23}{\Delta}_{21}} \tilde{V}_{\tau
3}\tilde{V}_{e 3}^{\ast} \ ,
\nonumber  \\
{V}_{\tau 3}{V}_{e 3}^{\ast} & = &
\frac{\widehat{\Delta}_{31}\widetilde\Delta_{23}}{{\Delta}_{31}
{\Delta}_{23}}\tilde{V}_{\tau 3}\tilde{V}_{e 3}^{\ast} +
\frac{\widehat{\Delta}_{33}
\widetilde\Delta_{21}}{{\Delta}_{31}{\Delta}_{23}} \tilde{V}_{\tau
1}\tilde{V}_{e 1}^{\ast} \
\end{eqnarray}
for $\triangle_\mu$; and
\begin{eqnarray}
{V}_{e1}{V}_{\mu 1}^{\ast} & = &
\frac{\widehat{\Delta}_{12}\widetilde\Delta_{31}}{{\Delta}_{12}
{\Delta}_{31}}\tilde{V}_{e1}\tilde{V}_{\mu 1}^{\ast} +
\frac{\widehat{\Delta}_{11}
\widetilde\Delta_{32}}{{\Delta}_{12}{\Delta}_{31}}
\tilde{V}_{e2}\tilde{V}_{\mu 2}^{\ast} \ ,
\nonumber  \\
{V}_{e2}{V}_{\mu 2}^{\ast} & = &
\frac{\widehat{\Delta}_{23}\widetilde\Delta_{21}}{{\Delta}_{23}
{\Delta}_{21}}\tilde{V}_{e2}\tilde{V}_{\mu 2}^{\ast} +
\frac{\widehat{\Delta}_{22}
\widetilde\Delta_{31}}{{\Delta}_{23}{\Delta}_{21}}
\tilde{V}_{e3}\tilde{V}_{\mu 3}^{\ast} \ ,
\nonumber \\
{V}_{e3}{V}_{\mu 3}^{\ast} & = &
\frac{\widehat{\Delta}_{31}\widetilde\Delta_{23}}{{\Delta}_{31}
{\Delta}_{23}}\tilde{V}_{e3}\tilde{V}_{\mu 3}^{\ast} +
\frac{\widehat{\Delta}_{33}
\widetilde\Delta_{21}}{{\Delta}_{31}{\Delta}_{23}}
\tilde{V}_{e1}\tilde{V}_{\mu 1}^{\ast} \
\end{eqnarray}
for $\triangle_\tau$, where $\Delta_{ij} \equiv m^2_i - m^2_j$
denote the neutrino mass-squared differences in vacuum. These
results are exactly the same as those obtained in Ref.
\cite{Zhang05}, although a different approach has been followed.
Eq. (22), (23) or (24) allows us to establish a direct relation
between $J$ and $\tilde{J}$ defined in Eq. (2). A straightforward
calculation yields $\tilde{J} \widetilde{\Delta}_{21}
\widetilde{\Delta}_{31} \widetilde{\Delta}_{32} = J \Delta_{21}
\Delta_{31} \Delta_{32}$, which has been derived in Refs.
\cite{SUM,Naumov} in a different way.

For the sake of completeness, let us list the expressions of two
independent ${\Delta}_{ij}$ in terms of their effective
counterparts in matter.
\begin{eqnarray}
{\Delta}_{31} & = & \displaystyle \frac{1}{3} \sqrt{\tilde{x}^2 -
3\tilde{y}} \left [3 \tilde{z} + \sqrt{3 \left (1 - \tilde{z}^2
\right )} \right ] \; ,
\nonumber \\
{\Delta}_{32} & = & \displaystyle \frac{1}{3} \sqrt{\tilde{x}^2 -
3\tilde{y}} \left [3 \tilde{z} - \sqrt{3 \left (1 - \tilde{z}^2
\right )} \right ] \; ,
\end{eqnarray}
where \cite{Xing01}
\begin{eqnarray}
\tilde{x} & = & \widetilde\Delta_{21} + \widetilde\Delta_{31} - A
\; ,
\nonumber \\
\tilde{y} & = & \displaystyle \widetilde\Delta_{21}
\widetilde\Delta_{31} - A \left [ \widetilde\Delta_{21} \left ( 1
- |\tilde{V}_{e2}|^2 \right ) + \widetilde\Delta_{31} \left ( 1 -
|\tilde{V}_{e3}|^2 \right ) \right ] \; ,
\nonumber \\
\tilde{z} & = & \displaystyle \cos \left [ \frac{1}{3} \arccos
\frac{2\tilde{x}^3 -9\tilde{x}\tilde{y} - 27 A
\widetilde\Delta_{21} \widetilde\Delta_{31} |\tilde{V}_{e1}|^2}{2
\left (\tilde{x}^2 - 3\tilde{y} \right )^{3/2}}\right ] \; .
\end{eqnarray}
Three independent $\widehat{\Delta}_{ij}$ can be expressed as
\begin{eqnarray}
\widehat{\Delta}_{11} & = & \displaystyle \frac{1}{3} \tilde{x} -
\frac{1}{3} \sqrt{\tilde{x}^2 - 3\tilde{y}} \left [ \tilde{z} +
\sqrt{3 \left (1 - \tilde{z}^2 \right )} \right ] \; ,
\nonumber \\
\widehat{\Delta}_{22} & = & \displaystyle  \frac{1}{3} \tilde{x} -
\frac{1}{3} \sqrt{\tilde{x}^2 - 3\tilde{y}} \left [ \tilde{z} -
\sqrt{3 \left (1 - \tilde{z}^2 \right )} \right ] -
\widetilde\Delta_{21} \; ,
\nonumber \\
\widehat{\Delta}_{33} & = & \displaystyle  \frac{1}{3} \tilde{x} +
\frac{2}{3} \tilde{z} \sqrt{\tilde{x}^2 - 3\tilde{y}} ~ -
\widetilde\Delta_{31} \; .
\end{eqnarray}
Note that $\widehat{\Delta}_{ij} = \Delta_{ij} +
\widehat{\Delta}_{jj} = \widetilde{\Delta}_{ij} +
\widehat{\Delta}_{ii}$ holds. Thus both ${\Delta}_{ij}$ and
$\widehat{\Delta}_{ij}$ are fully calculable, once
$\widetilde\Delta_{21}$, $\widetilde\Delta_{31}$, $A$,
$|\tilde{V}_{e1}|$ and $|\tilde{V}_{e2}|$ (or $|\tilde{V}_{e3}|$)
are specified.

Note that the afore-obtained results are only valid for neutrinos
propagating in vacuum and interacting with matter. As for
antineutrinos, the corresponding results can simply be obtained
through the replacements $V \Longrightarrow V^*$ and $A
\Longrightarrow -A$ (and $A' \Longrightarrow -A'$ for the
four-neutrino mixing scheme).

\section{Long-baseline neutrino oscillations}

The matter-corrected moduli $|\tilde{V}_{\alpha i}|^2$ and the
effective Jarlskog parameter $\tilde{J}$ can, at least in
principle, be determined from a variety of long-baseline neutrino
oscillation experiments. To be concrete, the survival probability
of a neutrino $\nu_\alpha$ and its conversion probability into
another neutrino $\nu^{~}_\beta$ are given by \cite{Review}
\begin{eqnarray}
\tilde{P}(\nu_{\alpha}\rightarrow\nu_{\alpha}) & = & 1 - 4
\sum\limits_{i<j} \left (|\tilde{V}_{\alpha i} \tilde{V}_{\alpha
j}^*|^2 \sin^2 \tilde{F}_{ji} \right ) \; ,
\nonumber \\
\tilde{P}(\nu_{\alpha}\rightarrow\nu^{~}_{\beta}) & = & - 4
\sum\limits_{i<j} \left [{\rm Re} \left (\tilde{V}_{\alpha i}
\tilde{V}_{\beta j} \tilde{V}_{\alpha j}^* \tilde{V}_{\beta i}^*
\right ) \sin^2 \tilde{F}_{ji} \right ] - 8\tilde{J} \prod
\limits_{i<j} \left(\sin \tilde{F}_{ji} \right ) \ ,
\end{eqnarray}
where $(\alpha, \beta)$ run over $(e,\mu)$, $(\mu,\tau)$ or
$(\tau,e)$, $\tilde{F}_{ji} \equiv 1.27
\widetilde{\Delta}_{ji}L/E$ with $L$ being the baseline length (in
unit of km) and $E$ being the neutrino beam energy (in unit of
GeV), and $\tilde{J}$ has been defined in Eq. (2). The ${\rm Re}
(\tilde{V}_{\alpha i} \tilde{V}_{\beta j} \tilde{V}_{\alpha j}^*
\tilde{V}_{\beta i}^*)$ term in
$\tilde{P}(\nu_{\alpha}\rightarrow\nu^{~}_{\beta})$ can be
expressed as
\begin{equation}
{\rm Re} \left (\tilde{V}_{\alpha i} \tilde{V}_{\beta j}
\tilde{V}_{\alpha j}^* \tilde{V}_{\beta i}^* \right ) =
\frac{1}{2} \left (|\tilde{V}_{\alpha k} \tilde{V}^*_{\beta k}|^2
- |\tilde{V}_{\alpha i} \tilde{V}^*_{\beta i}|^2 -
|\tilde{V}_{\alpha j} \tilde{V}^*_{\beta j}|^2 \right ) \;
\end{equation}
with $\alpha \neq \beta$ and $i\neq j\neq k$. Then we rewrite Eq.
(28) as follows:
\begin{eqnarray}
\tilde{P}(\nu_{\alpha}\rightarrow\nu_{\alpha}) & = & 1 - 4 \left
(|\tilde{V}_{\alpha 1} \tilde{V}_{\alpha 2}^*|^2 \sin^2
\tilde{F}_{21} + |\tilde{V}_{\alpha 1} \tilde{V}_{\alpha 3}^*|^2
\sin^2 \tilde{F}_{31} + |\tilde{V}_{\alpha 2} \tilde{V}_{\alpha
3}^*|^2 \sin^2 \tilde{F}_{32} \right ) \; ,
\nonumber \\
\tilde{P}(\nu_{\alpha}\rightarrow\nu^{~}_{\beta}) & = & - 2 \left
[ |\tilde{V}_{\alpha 1} \tilde{V}^*_{\beta 1}|^2 S_1 +
|\tilde{V}_{\alpha 2} \tilde{V}^*_{\beta 2}|^2 S_2 +
|\tilde{V}_{\alpha 3} \tilde{V}^*_{\beta 3}|^2 S_3 \right ] - 8
\tilde{J} S^{~}_J \; ,
\end{eqnarray}
where
\begin{eqnarray}
S_1 & \equiv & \sin^2\tilde{F}_{32} - \sin^2\tilde{F}_{21} -
\sin^2\tilde{F}_{31} \; ,
\nonumber \\
S_2 & \equiv & \sin^2\tilde{F}_{31} - \sin^2\tilde{F}_{21} -
\sin^2\tilde{F}_{32} \; ,
\nonumber \\
S_3 & \equiv & \sin^2\tilde{F}_{21} - \sin^2\tilde{F}_{31} -
\sin^2\tilde{F}_{32} \; ,
\nonumber \\
S^{~}_J & \equiv & \sin\tilde{F}_{21} \sin\tilde{F}_{31}
\sin\tilde{F}_{32} \; .
\end{eqnarray}
Some comments on Eqs. (30) and (31) are in order.
\begin{itemize}
\item  Three oscillation terms in $\tilde{P}(\nu_\alpha
\rightarrow \nu_\alpha)$ are associated with $|\tilde{V}_{\alpha
i}\tilde{V}^*_{\alpha j}|^2$ (with $i\neq j$). Taking $i=1$ and
$j=2$ for example, we obtain an effective unitarity triangle whose
three sides are $\tilde{V}_{e 1}\tilde{V}^*_{e 2}$,
$\tilde{V}_{\mu 1}\tilde{V}^*_{\mu 2}$ and $\tilde{V}_{\tau
1}\tilde{V}^*_{\tau 2}$ in the complex plane. This triangle was
originally named as $\widetilde{\triangle}_3$ in Ref. \cite{FX}.
Triangles $\widetilde{\triangle}_1$ (for $i=2$ and $j=3$) and
$\widetilde{\triangle}_2$ (for $i=1$ and $j=3$) can similarly be
defined. Their genuine counterparts in vacuum are referred to as
$\triangle_i$ for $i=1,2$ and 3. If the sides of
$\widetilde{\triangle}_i$ can all be determined from
$\tilde{P}(\nu_\alpha \rightarrow \nu_\alpha)$ in some {\it
disappearance} neutrino oscillation experiments, it is then
possible to calculate individual $|\tilde{V}_{\alpha i}|^2$ and to
extract $|V_{\alpha i}|^2$ by using Eqs. (17), (18) and (19).
\item  Three sides of triangle $\widetilde{\triangle}_e$,
$\widetilde{\triangle}_\mu$ or $\widetilde{\triangle}_\tau$ are
associated with three oscillation terms $S_i$ (for $i=1,2,3$) of
the {\it appearance} neutrino oscillation probability
$\tilde{P}(\nu_\alpha \rightarrow \nu_\beta)$, while the effective
CP-violating parameter $\tilde{J}$ is relevant to the oscillation
term $S_J$. To illustrate, we plot the dependence of $S_i$ and
$S_J$ on the neutrino (or antineutrino) beam energy $E$ in Fig. 1,
where two typical neutrino baselines $L=730 ~{\rm km}$ \cite{Long}
and $L=2100 ~{\rm km}$ \cite{H2B} have been taken. The input
parameters include $\Delta_{21} \approx 8\times 10^{-5} ~ {\rm
eV}^2$, $\Delta_{32} \approx 2.3\times 10^{-3} ~ {\rm eV}^2$,
$\theta_{12} \approx 33^\circ$, $\theta_{23} \approx 45^\circ$,
$\theta_{13} \approx 3^\circ$ and $\delta \approx 90^\circ$ in the
standard parametrization of $V$ \cite{Review}. In addition, the
terrestrial matter effects can approximately be described by $A
\approx 2.28\times 10^{-4} {\rm eV}^2 E/[{\rm GeV}]$
\cite{Shrock}. Fig. 1 shows that $S_i$ may have quite different
behaviors, if the baseline length is sufficiently large (e.g., $L
\sim 1000$ km or larger). Hence the proper changes of $E$ and (or)
$L$ would allow us to determine the coefficients of $S_i$, i.e.,
$|\tilde{V}_{\alpha i}\tilde{V}^*_{\beta i}|^2$. The parameter
$\tilde{J}$ can also be extracted from a suitable long-baseline
neutrino oscillation experiment, because the dependence of $S_J$
on $E$ and $L$ is essentially different from that of $S_i$.
Provided all or most of such measurements are realistically done,
the moduli of $V_{\alpha i}$ and leptonic unitarity triangles
$\triangle_e$, $\triangle_\mu$ and $\triangle_\tau$ may finally be
reconstructed. \item  It is clear that both types of neutrino
oscillation experiments (i.e., appearance and disappearance) are
needed, in order to get more information on lepton flavor mixing
and CP violation. They are actually complementary to each other in
determining the moduli of nine matrix elements of $V$ and its
unitarity triangles. In practice, the full reconstruction of $V$
from $\tilde{V}$ requires highly precise and challenging
measurements. A detailed analysis of the unitarity triangle
reconstruction can be found from Ref. \cite{Smirnov}, in which the
issues of experimental feasibility and difficulties have more or
less been addressed.
\end{itemize}

Let us remark that the strategy of this paper is to establish the
model-independent relations between $V$ and $\tilde{V}$, both
their moduli and their unitarity triangles. Hence we have
concentrated on the generic formalism instead of the specific
scenarios or numerical analyses. Our exact analytical results are
expected to be a very useful addition to the phenomenology of
lepton flavor mixing and neutrino oscillations.

\section{Summary}

We have derived a new set of sum rules between $(m_i, V)$ and
$(\tilde{m}_i, \tilde{V})$ for both three- and four-neutrino
mixing schemes. With the help of our sum rules, we have calculated
the moduli of $V_{\alpha i}$ in terms of those of
$\tilde{V}_{\alpha i}$. The sides of leptonic unitarity triangles
$\triangle_e$, $\triangle_\mu$ and $\triangle_\tau$ can also be
figured out in a similar way. We find that it is useful to express
the neutrino oscillation probabilities in terms of
$|\tilde{V}_{\alpha i}|^2$ and the matter-corrected Jarlskog
parameter $\tilde{J}$. The fundamental lepton flavor mixing matrix
$V$ and its six unitarity triangles can then be reconstructed, at
least in principle, from a variety of long-baseline neutrino
oscillations in a straightforward and parametrization-independent
way.

Our analytically exact results will be applicable for the study of
terrestrial long-baseline neutrino oscillations, from which much
better understanding of the neutrino mass spectrum, lepton flavor
mixing and CP violation can be achieved.

\vspace{0.5cm}

One of us (Z.Z.X.) is grateful to CSSM at the University of
Adelaide, where this paper was written, for warm hospitality. Our
work was supported in part by the National Nature Science
Foundation of China.

\begin{figure}
\begin{center}
\vspace{-2cm}
\includegraphics[width=16cm,height=24cm]{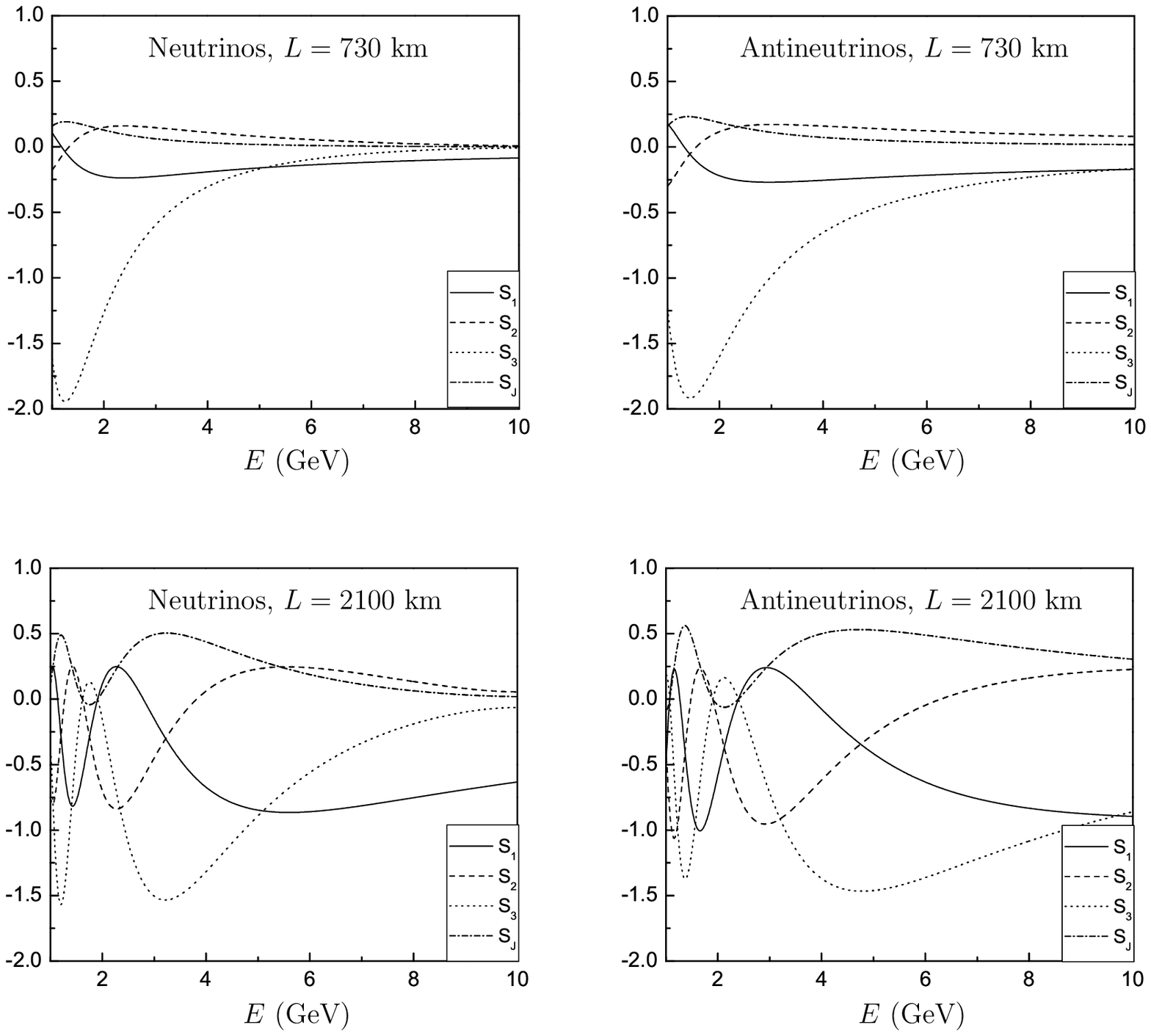}
\vspace{-7cm} \caption{Numerical illustration of four different
oscillation terms appearing in $\tilde{P}(\nu_\alpha \rightarrow
\nu_\beta)$ and $\tilde{P}(\overline{\nu}_\alpha \rightarrow
\overline{\nu}_\beta)$, where two typical neutrino beam baselines
have been taken.}
\end{center}
\end{figure}

\end{document}